# SOI Pixel Sensor for Gamma-Ray Imaging

KENJI SHIMAZOE, FAIRUZ ATIQAH, YURI YOSHIHARA, AKIHIKO KOYAMA, HIROYUKI TAKAHASHI, TADASHI ORITA, KEI KAMADA, AYAKI TAKEDA, TAKESHI TSURU, YASUO ARAI

[1]The University of Tokyo, Tokyo, JAPAN
[2]Japan Atomic Energy Agency, Ibaraki, JAPAN
[3]Tohoku University, Sendai, JAPAN
[4]Kyoto University, Kyoto, JAPAN
[5]KEK, Tsukuba, JAPAN
[6]OIST, Okinawa, JAPAN

The application of SOI pixel sensor to gamma-ray imaging is described using electron tracking Compton imaging.

PRESENTED AT

# 1 Introduction

Compton Imaging is very useful imaging method to localize the radioactive source emitting sub MeV to a few MeV gamma-rays because of the higher probability of Compton scattering compared with photo absorption event in the energy range. Compton imaging has many application field, such as environmental radiation monitoring, gamma-ray astronomy, and medical imaging. One of the typical target radiation energy is 662 keV from radioactive $^{137}$Cs in the environmental monitoring and other is 511 keV emitted from positron emitting radionuclides used in the medical applications. There has been developed several prototypes of Compton imager based on different type of materials, semiconductors and scintillation crystals. These are using two measured energies in the scatters and absorbers to make the final reconstructed image using full Compton cones [1]-[4]. There has been also investigated new imaging system called advanced Compton Imaging (advanced Compton camera ACC) by a few groups for utilizing the direction of recoiled electrons by Compton scattering and partial Compton cones [5] [6]. It is reported ACC could improve the total sensitivity up to 5 - 10 times, but the currently developed systems are based on gaseous detector and charge coupled device (CCD) [7]. It is not yet reported the system with solid state detectors with the practical readout speed (~k Hz) utilizing trigger signals. Typical Compton imaging system requires the speed of from k Hz to MHz readout and its coincidence detection function.

SOI (Silicon-On-Insulator) pixel sensor [8] is promising technology for developing the high position resolution detector by integrating the small pixels and circuits in the monolithic way. The event driven (trigger mode) SOI based pixel sensor has also been developed for the application of X-ray astronomy with the purpose of reducing the noise using anti-coincidence event [9]. This trigger mode SOI pixel sensor working with in the rate of kilo Hz is also a promising scatter detector for advanced Compton imaging to track the Compton recoiled electrons.

# 2 Materials and Methods

The trigger mode SOI pixel sensor used in the experiment has a 144 × 144 effective pixel array with the pixel size of 30 μm [9]. The effective area is 4.3 mm × 4.3 mm with the thickness of 250 μm (CZ) or 500 μm (FZ). The internal core circuit works with the voltage of 1.8 V and IO works with 3.3V. The spectrum is measured with $^{241}$Am and $^{109}$Cd radiation source with room temperature and the back bias voltage of 5 V.

The track of recoiled electrons is observed using frame mode and trigger mode with room temperature and the back bias voltage of 5 V. The pattern readout with from 3 × 3 to 25 × 25 pixels surrounding the triggered pixel is implemented in the program on SEABAS (Soi EvAluation Board with SiTCP) [10] based FPGA data acquisition system for realizing the faster readout compared with the frame readout of 144 × 144 pixels. The pattern readout is required to track the full trajectory of Compton recoiled electrons, which corresponds to the size up to 750 μm × 750 μm. The SOI sensor was irradiated with $^{60}$Co and $^{241}$Am source.

48 × 48 GAGG pixel array is fabricated with the pitch of 500 μm and the crystal size of 0.4 × 0.4 × 20 mm$^2$ as an absorber of Compton imaging system. Figure. 2 shows the picture of GAGG array wrapped with teflon tape. Each crystal is separated with BaSO4 based reflector

The coincidence system between the trigger mode SOI pixel sensor and scintillation crystal coupled to SiPM is fabricated to form the Compton imaging system. The coincidence events are recorded using the FPGA based DAQ and a preliminary image is measured with the experiment using $^{137}$Cs source.

# 3 Results

Figure 1 shows the measured spectra of one pixel with $^{241}$Am and $^{109}$Cd source using SOI pixel sensor (CZ-type) with trigger mode configuration. The threshold for the trigger is set 450 mV in this experiment. The energy resolution of 1.2 keV is observed at 17.6 keV peak. Good linearity is observed in the range of from 10 keV to 60 keV.

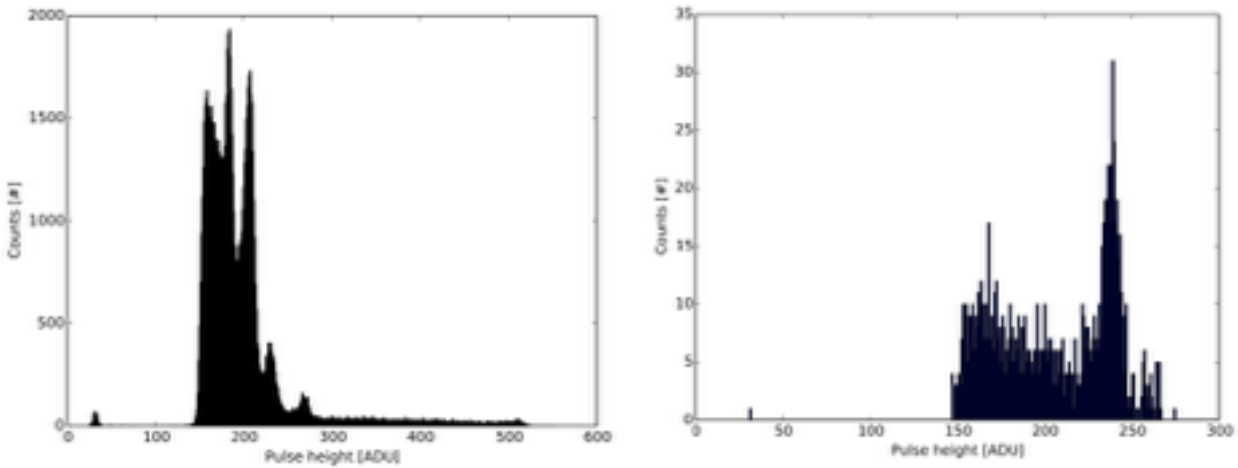

Figure 1. energy spectra with $^{241}$Am (left) and $^{109}$Cd (right) of trigger mode SOI sensor (CZ-type)

Figure 2 shows the captured trajectory of recoiled electrons irradiated with $^{60}$Co source and $^{241}$Am source. The SOI pixel sensor (CZ-type) is operated with the frame mode at the room temperature in this setup. Left panel in the figure shows the typical trajectory of electrons irradiated by $^{60}$Co and its magnified image. The track reaches to the length of approximately more than 300 μm. Right panel shows the energy deposition by $^{241}$Am source and the charges are typically shared within 2 or 3 pixels. These images are measured with the integration time of 1000 μs.

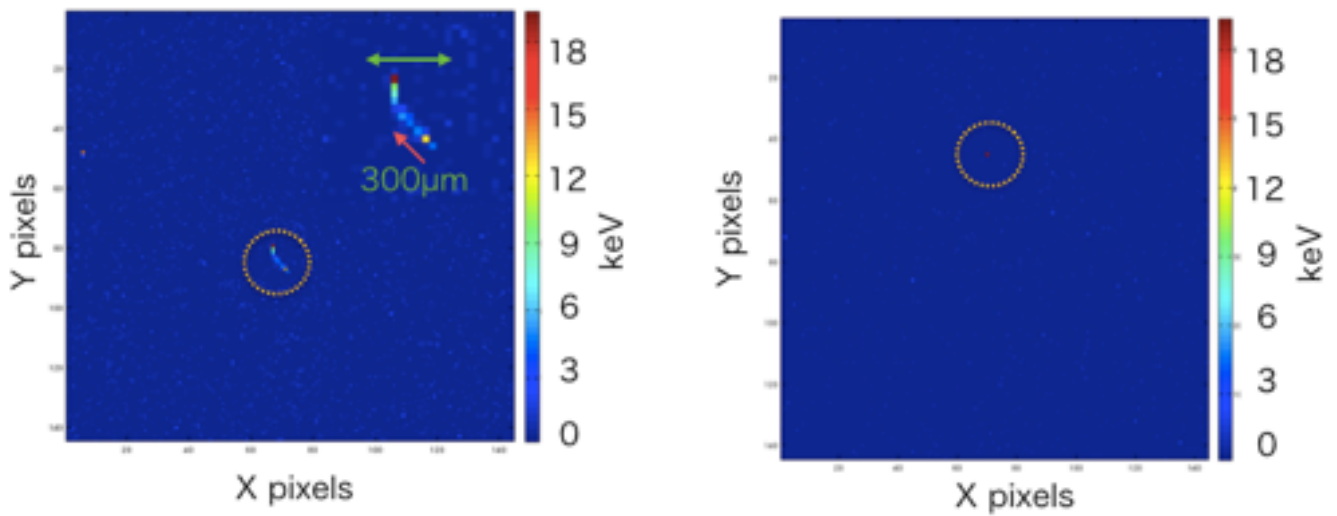

Figure 2. captured trajectory of recoiled electrons irradiated with $^{60}$Co (left) and $^{241}$Am (right)

The trigger mode pattern readout is implemented in the FPGA program to read out the energy deposition surrounding the triggered pixel. Figure 3 shows the typical trajectory of electrons within 25 × 25 pixels irradiated with $^{60}$Co source. The trajectory of recoil electron is clearly observed although there observed some noises caused by the cross-talk between digital and analog circuit. It is thought the slightly higher energy deposition at the lowest column is caused by the cross-talk noise.

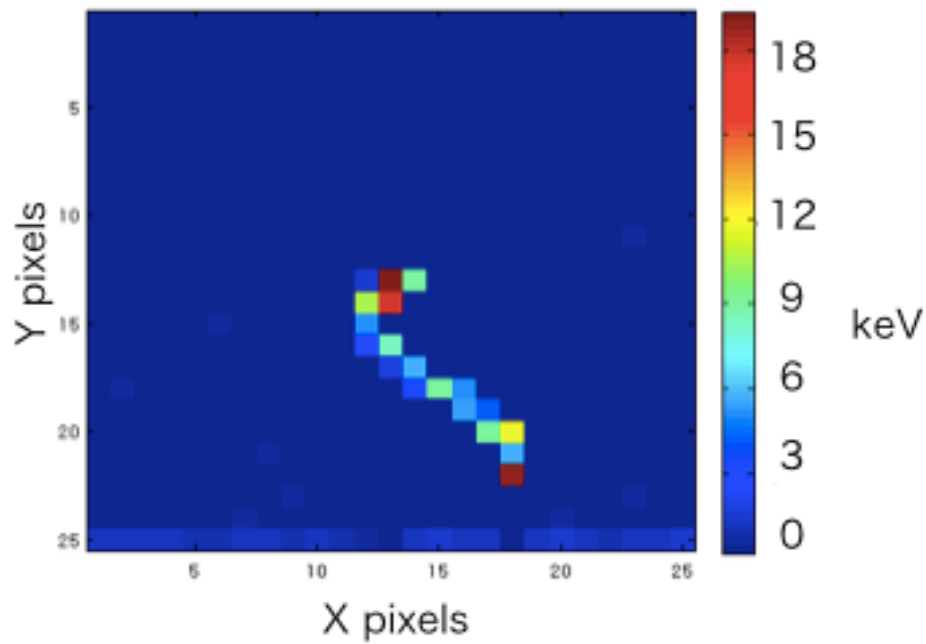

Figure 3. typical trajectory of recoil electrons using 25 × 25 pixels pattern readout in trigger mode SOI pixel sensor

Figure 4 shows the picture of fabricated 48 × 48 GAGG pixel array detector for absorber wrapped with teflon tape. One pixel consists of 0.4 × 0.4 × 20 mm$^3$ GAGG crystals separated with 0.1 mm BaSO$_4$ reflector. The flood map of the array was measured by coupling GAGG array to MPPC (TSV-type S12642-0808PA-50). The signals were read out with charge division method using 4 channels using resistive chain. The each pixel was separated clearly except the channels on the edge.

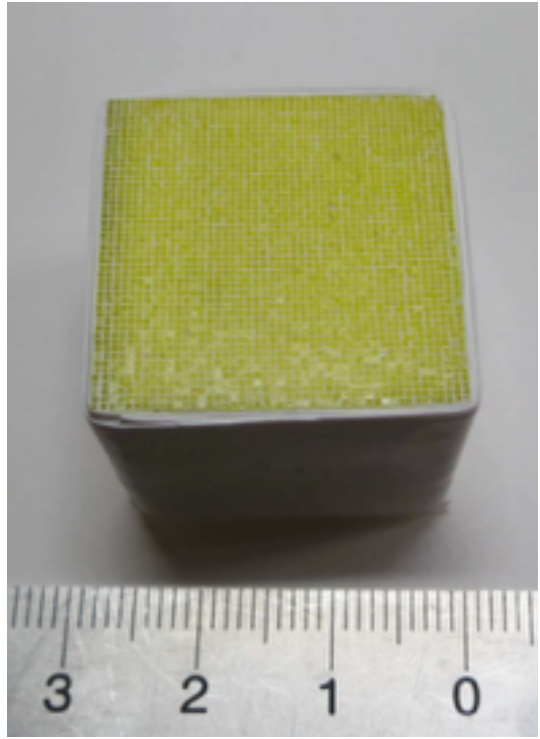

Figure 4.  48 × 48 GAGG pixel array with 0.4 × 0.4 × 20 mm$^3$ GAGG crystals

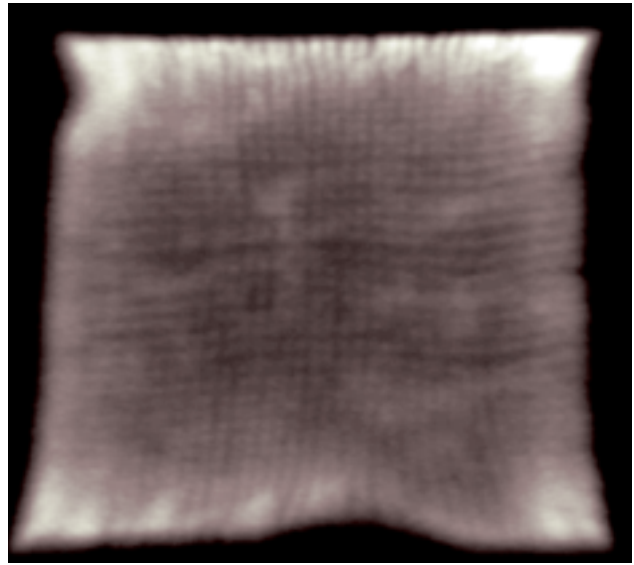

Figure 5.  48 × 48 GAGG pixel array with 0.4 × 0.4 × 20 mm$^3$ GAGG crystals

# 4 Conclusion

Aiming the solid state advanced Compton imaging, the capability of electron tracking in SOI pixel sensor was investigated. The track of Compton recoiled electrons was observed with the accuracy of 30 μm position resolution using the trigger mode SOI pixel sensor. The pattern readout with 3 × 3 and 25 × 25 pixels surrounding the triggered pixel was implemented to track the full trajectory of recoiled electrons.  The coincidence system between the trigger mode SOI pixel sensor and the scintillator and SiPM based absorber was fabricated for the future Compton imaging system. Three dimensional reconstruction of recoiled electron trajectory and the Compton imaging has to be investigated in the successive work.